\documentclass[sigconf]{acmart}

\def\BibTeX{{\rm B\kern-.05em{\sc i\kern-.025em b}\kern-.08emT\kern-.1667em\lower.7ex\hbox{E}\kern-.125emX}}

\usepackage{wrapfig}
\usepackage{enumitem}
\usepackage{soul}
\usepackage{subfigure}

\usepackage{graphbox}

\newcommand{\mm}[1]{{\textcolor{black}{#1}}}

\setcopyright{acmlicensed}
\copyrightyear{2019}
\acmYear{2019}
\acmConference[DocEng '19]{ACM Symposium on Document Engineering 2019}{September 23--26, 2019}{Berlin, Germany}
\acmBooktitle{ACM Symposium on Document Engineering 2019 (DocEng '19), September 23--26, 2019, Berlin, Germany}
\acmPrice{15.00}
\acmDOI{10.1145/3342558.3345395}
\acmISBN{978-1-4503-6887-2/19/09}

\begin{document}
	
	%
	\title{Augmenting Music Sheets with Harmonic Fingerprints}
	
	%

	\author{Matthias Miller}
	\affiliation{%
		\institution{Department of Computer Science}
		\streetaddress{Universit\"atsstrasse 10}
		\city{University of Konstanz}
		\country{Germany}}
	\email{matthias.miller@uni.kn}
	
	\author{Alexandra Bonnici}
	\affiliation{%
		\institution{Faculty of Engineering}
		\streetaddress{}
		\city{University of Malta}
		\country{Malta}}
	\email{alexandra.bonnici@um.edu.mt}

	\author{Mennatallah El-Assady}
	\affiliation{%
		\institution{Department of Computer Science}
		\streetaddress{Universit\"atsstrasse 10}
		\city{University of Konstanz}
		\country{Germany}}
	\email{menna.el-assady@uni.kn}
	
	%
	\renewcommand{\shortauthors}{Miller, Bonnici, and El-Assady.}

	\begin{teaserfigure}
		\begin{center}
			\includegraphics[width=\textwidth]{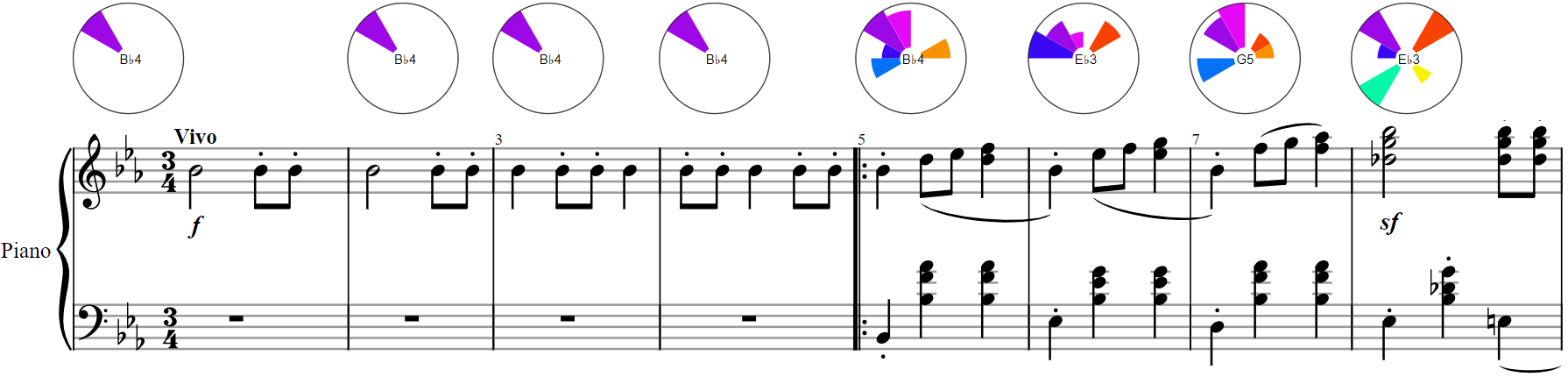}
		\end{center}
		\caption{Augmenting piano sheet music with \textit{harmonic fingerprint glyphs} facilitates the identification of recurring harmonic patterns and the comparison of musical parts to understand differences in the note distribution. Here, an excerpt from Chopin's `\textit{Grande Valse Brillante}' is augmented with the fingerprints showing a recurring pattern in the first four glyphs.}
		\Description{.}
		\label{fig:teaser} 
	\end{teaserfigure}

	\begin{abstract}
			Common Music Notation (CMN) is the well-established foundation for the written communication of musical information, such as rhythm or harmony. CMN suffers from the complexity of its visual encoding and the need for extensive training to acquire proficiency and legibility. 
			While alternative notations using additional visual variables (e.g., color to improve pitch identification) have been proposed, the community does not readily accept notation systems that vary widely from the CMN. 
			Therefore, to support student musicians in understanding harmonic relationships, instead of replacing the CMN, we present a visualization technique that augments digital sheet music with a \textit{harmonic fingerprint glyph}. Our design exploits the circle of fifths, a fundamental concept in music theory, as visual metaphor.
			By attaching such glyphs to each bar of a composition we provide additional information about the salient harmonic features available in a musical piece. We conducted a user study to analyze the performance of experts and non-experts in an identification and comparison task of recurring patterns. The evaluation shows that the harmonic fingerprint supports these tasks without the need for close-reading, as when compared to a not-annotated music sheet. 
	\end{abstract}


	
	\maketitle
	
	\setlength{\columnsep}{5pt}
	
	\section{Introduction}
	Isidore of Seville recounts how until the 7th-century music was only conserved as an auditory memory as there was yet no means of notating music~\cite{barney_etymologies_2006}. 
	By the 9th century, neumes came into use 
	as visual aids to indicate the relative pitch and melody direction in  Gregorian chant~\cite{strayer2013neumes}. 
	Over several centuries this neumatic notation was adapted to reflect the evolution of musical instruments as well as different music composition styles, 
	to create the Conventional Music Notation (CMN) used today. 
	Despite the widespread use of the CMN, the notation faces criticism on two counts namely that the notation is not adequate to represent musical notation, and that it does not adequately represent all information of a musical score. 
	The CMN is thought to be inadequate as a musical notation because it is mainly based on the use of a five-line staff such that only nine vertical spaces are available to represent the 12 tones of the chromatic scale. Thus the CMN introduces sharp and flat signs to represent all 12 tones, unnecessarily complicating the visualization of the note pitches. Likewise, the notation of the note durations does not intuitively represent the score's rhythmic properties.  
	
	These criticisms gave rise to alternative music notation systems to facilitate the reading of music, especially for novice learners. Among these alternative systems, we may find notation systems such as Klavarskribo~\cite{kaasjager_2001_klavarscribo}, Hummingbird~\cite{humnotation},  Dodeka~\cite{dodekanotation} and Pizzicato~\cite{pizzicatonotation} among others. These alternative notation systems may include the introduction of different staff systems to capture equally spaced chromatic intervals~\cite{kaasjager_2001_klavarscribo}, the use of different symbols to represent note pitches and durations~\cite{humnotation, kuo_proposal_2013}, piano-roll style music representations~\cite{dodekanotation}, and numeric-based notations~\cite{pizzicatonotation}.
	
	However, these alternative notations have not been adopted by the music community due to the lack of accessibility of music scores written in the different notation and because the number of musicians who can read the CMN by far outnumbers those who can read alternative notations. Thus, the problem of reading and quickly understanding the CMN remains a struggle for novices~\cite{malandrino_visualization_2018}.  
	
	Technology, notably, screen applications have changed the way musicians are sharing and using musical scores~\cite{Sbastien2013AnnotatingWF}. Music scores are no longer restricted to print media. Advances in optical music recognition as well as the automated conversion of musical scores into different file formats led to the emergence of alternative notations. Instead of providing a complete overhaul of the CMN, often additional score annotations are employed as additional teaching aids~\cite{de_prisco_understanding_2017}. Furthermore, the digital representation of music as web or tablet-based apps makes it possible to have both conventional representations and new notations available for the same document such that the student may switch between notation systems easily. 
	
	In general, sheet music encodes musical features such as rhythm, dynamics, harmony, and other instructions as visual symbols which specify how a performer can reproduce a composition~\cite{miller_analyzing_2018}. The notation, however, omits the explicit representation of the relationship that exists between the individual notes and their location within the overarching musical structure~\cite{hunt2017can}. While understanding these complex relationships is of immediate interest for the musician~\cite{de_haas_automatic_2013}, such an understanding typically requires analyzing the score and thus, a certain level of musical expertise~\cite{malandrino_visualization_2018, de_prisco_understanding_2017}. The understanding of the harmonic structure of the score allows for faster learning of new musical pieces and enhances musical memory and should ideally be more accessible to novices.  Annotating the score with a visual representation of the hidden harmonic patterns would, therefore, help to convey this musical knowledge leading to a deeper understanding of the underlying musical structure~\cite{malandrino_visualization_2018}. 
	
	In this paper, we contribute a \textit{harmonic fingerprint glyph} visualization to encode and augment the distribution of notes in pre-defined windows within musical pieces. This avoids the introduction of new music notations that differm from the conventional music notation. We use the circle of fifths~\cite{heinichen_generalbass_1969} as a visual metaphor to design the harmonic fingerprint. By attaching this glyph, we add valuable information about the harmonic content to digital sheet music, giving readers the opportunity to \textit{efficiently} identify and compare harmonic relationships across a document. To test the performance of the fingerprint, we conducted a qualitative user study, collecting comparative feedback from both domain experts and non-experts. We concentrate on the following three research questions to emphasize the benefits and drawbacks of our approach:
	\begin{enumerate}
		\item[\textbf{Q1}] Does the proposed fingerprint annotation method support the \textit{identification} of recurring and long musical patterns? 
		\item[\textbf{Q2}] Does the fingerprint annotation influence the harmony \textit{analysis} of the score?
		\item[\textbf{Q3}] Does the reader's musical \textit{expertise} level influence the effectiveness of the fingerprint for performing harmony analysis?
	\end{enumerate}
	
	\section{Related Work}
	\label{background}
	\mm{Understanding harmonic relationships and tonal progressions of a musical composition is a fundamental aspect of music analysis~\cite{sapp_visual_2005}. To reveal these connections of harmony, musicians need to analyze the music to identify the chords use and identify any patterns using knowledge of music theory.  Performing such a task requires knowledge and expertise, which are not necessarily within reach of beginner musicians~\cite{malandrino_visualization_2018}. Harmony visualization tools make harmonic analysis more accessible by providing visual aids which highlight non-explicit information of the score. Visualization algorithms typically follow Shneiderman's \emph{Visual Information Seeking Mantra}, that is, follow the three rules: overview first; zoom and filter; then details-on-demand~\cite{shneiderman_eyes_1996}}.
	
	\subsection{Music Harmony Visualization}
	There are different ways to visualize harmonic relationships in musical works. Abstract music visualizations enable obtaining an overview even of full musical pieces, but readers are not able to look at the details anymore. In contrast, providing original details of music score facilitate the understanding of the exact harmonic and rhythmic relationships on a lower level, but the reader can not readily recognize the overall structure.  \citeauthor{jaenicke_closedistantreading_2015} explain that combining close and distant reading allows users to find regions of interest while still providing all details instead of only presenting an abstract representation. In this way, readers are directed by distant reading while investigating the underlying details afterward~\cite{jaenicke_closedistantreading_2015}. We transfer \citeauthor{jaenicke_closedistantreading_2015}'s close and distant reading concept from the text domain to music visualization by showing its importance for the visualization harmony in music~\cite{jaenicke_closedistantreading_2015}.
	
	\subsubsection{Distant Reading}\ \\
	Analyzing particular attributes of structural data and providing them in a visual way is a typical approach to reveal unknown patterns. For example, \citeauthor{keim_literature_2007} present a method to apply visual, literary analysis for different types of features that are characteristic of text~\cite{keim_literature_2007}. The proposed visual literature fingerprinting method has proven to be an effective approach to show meaningful differences between several text sections while including various text features such as average sentence length for creating the fingerprint visualization. \citeauthor{keim_literature_2007} do not include close reading, so users who want to further investigate the details in interesting areas of the visualization are not supported. 
	
	One approach to visualize the harmonic content of a musical piece is to create a representation which displays the overall form of the music. For example, \citeauthor{wattenberg_arc_2002} introduces an \emph{arc diagram} visualization to reveal sequential melodic patterns. Sequential repetitions of melody notes are highlighted by arcs connected the notes. The arc's radius and width indicative of the distance and similarity between recurring melodic sequences~\cite{wattenberg_arc_2002}. Similar approaches include the representation of the score using \emph{similarity matrix} visualizations~\cite{wolkowicz_midivis_2009}, or the use of the Tonnetz grid to create an \emph{Isochord} visualization of the score~\cite{bergstrom_isochords_2007}. Using a similar approach, \citeauthor{schroer_visualizingharmony_2019} uses \emph{circos graphs} which are used to reveal patterns in genomic data. Using these graphs, \citeauthor{schroer_visualizingharmony_2019} represents the harmonic relationship between all the twelve tones of the chromatic scale by linking the chord root with other notes that are played simultaneously~\cite{schroer_visualizingharmony_2019}. 
	
	Such diagram representations of the score emphasize the commonalities that exist between whole sections of the musical piece. Therefore, these visualizations can highlight similarities in structure that exist between different musical scores that are typical for a genre. However, these approaches do not retain the relation to the original score notation. While it is easy to visualize the overall structure of the score, the spatial position of these highlighted structures gets lost. Enabling music readers to retrieve the detailed CMN on demand is essential to support understanding of the underlying harmonies. This requires to combine distant reading with close reading to exploit the advantages of both methods.

	\subsubsection{Close Reading}\ \\
	Local visualizations of the score retain the spatial information of the highlighted structure concerning the full score. \citeauthor{smith_visualization_1997} exploit three-dimensional space and color to visualize typical musical information based on MIDI files~\cite{smith_visualization_1997}. They propose to map tone data described by pitch, volume, and timbre to colored spheres in their visualization. The aim of \citeauthor{smith_visualization_1997} is to visually present music to listeners who are not familiar with reading music notation. Rather than providing a static visualization, their model uses a dynamic visual model which decreases the color intensity of the notes with the progress of time, thus allowing the listener to distinguish between different tones played at different points in time. While such an approach is visually pleasing, it does not allow the listener to understand the underlying harmonic sections or recurring patterns easily.
	
	Algorithms such as that described by \citeauthor{snydal_improviz_2005} create melodic landscapes and harmonic palettes from transcriptions of jazz improvisations~\cite{snydal_improviz_2005}, while \citeauthor{miyazaki_exploring_2003} use cylinders whose height, diameter and color represent pitch, volume and duration information of the notes in the score. In the latter algorithm, the visualization can be represented both at an overview level and also at a detailed level according to user preference~\cite{miyazaki_exploring_2003}. 
	\citeauthor{ciuha_visualization_2010} describe an approach to visualize concurrent tones, using color to highlight the harmony~\cite{ciuha_visualization_2010}. In this visualization, the horizontal axis represents the temporal progress of a single musical piece while the vertical axis encodes the note pitches. Depending on the temporal segment size the visualization is blurred based on the tonal unambiguousness. In this way, this method conveys the harmonic journey of a piece while using an appealing representation to encode the affinity of notes by applying a color wheel on the circle of fifths to reflect the dissonance of intervals.
	
	\citeauthor{sapp_visual_2005} proposes a visual method which emphasizes the key strength at each section~\cite{sapp_visual_2005}. This model uses color to encode the dominant notes or key at several hierarchical levels. At the top of the hierarchy, the model represents the most significant key of the entire composition. The lower levels of the hierarchy represent detailed tonal progressions, resulting in a triangular representation of the keys used in the music. Thus, this model is especially useful to reveal the tonal variations in the piece. 
	
	\subsection{Annotating the CMN Score}
	The visualization strategies discussed so far are detached from the CMN which makes it difficult to keep the connection between the original music sheets and the visualization. One strategy to resolve this problem is to display the original notation format and annotate the score. An example of such an approach commonly used in music notation is the placement of chord symbols to indicate the accompanying harmonies which the musician needs to perform. However, these chord symbols are a rather general description of the harmony progressions and cannot reflect the complexity of specific musical parts in detail. Thus, while scores can be annotated using a similar analogy, an annotated visualization would let the reader focus on the harmonic aspects of the score. 
	
	Addressing these concerns, \citeauthor{cuthbert_music21_2010} describe \texttt{\emph{music21}}, a tool which provides a number of features for visual score descriptions. Among these, the harmonic analysis of the score is shown as chords written in closed-form. The metric analysis is represented as asterisk signs beneath the notes, while a plot of the pitch class against the bar number illustrates the pitch usage over the progression of the score. 
	
	The \emph{VisualHarmony} tool described by \citeauthor{malandrino_visualization_2018} overlays visual information about the tonal information directly onto the score CMN. The tool aims to support music composition by highlighting score parts that do not comply with classical music theory rules~\cite{malandrino_visualization_2018}. The tool facilitates the identification of chord tonality and the respective scale degrees. This is used to display melodic errors supporting the music composition task. In VisualHarmony, the visual information is shown as colored rectangular boxes enclosing the individual chords. Thus, the system preserves the spatial location of the harmonic patterns. However, the overlapping visualization was found to be too distracting, particularly for directly playing the music and thus, \citeauthor{de_prisco_evaluation_2018} suggest that the visual information should be displayed above the staff~\cite{de_prisco_evaluation_2018}.
	
	From this overview of the related work, we may note that visualization algorithms which represent the melodic patterns in the music, do so on a global level, without preserving the spatial location of these patterns. While visualizations such that described by~\cite{malandrino_visualization_2018} do preserve the spatial location, these focus on analyzing the harmony of the music. While this is important, the harmonic analysis does not capture the melodic patterns that can also be present in the score. Thus, an annotation system which captures both the harmonic patterns and the melodic patterns would be of benefit to the community.

	\section{Harmony Fingerprint Design} 
	\label{methodology}
	\mm{
		This section describes the proposed harmonic fingerprint visualization. This visualization approach will capture the harmonic and melodic content of each bar of the musical score. It exploits the circle of fifths as a visual metaphor for its design. While highlighting the harmonic information contained within a bar, the glyph also represents tonal information in a way with which musicians may already be familiar.
	}
	
	\subsection{Circle of Fifths}
	
\begin{figure}[t]
		\subfigure[][The musical circle originally published by Heinichen in 1728~\cite{heinichen_generalbass_1969}]{\includegraphics[width = 0.49\linewidth]{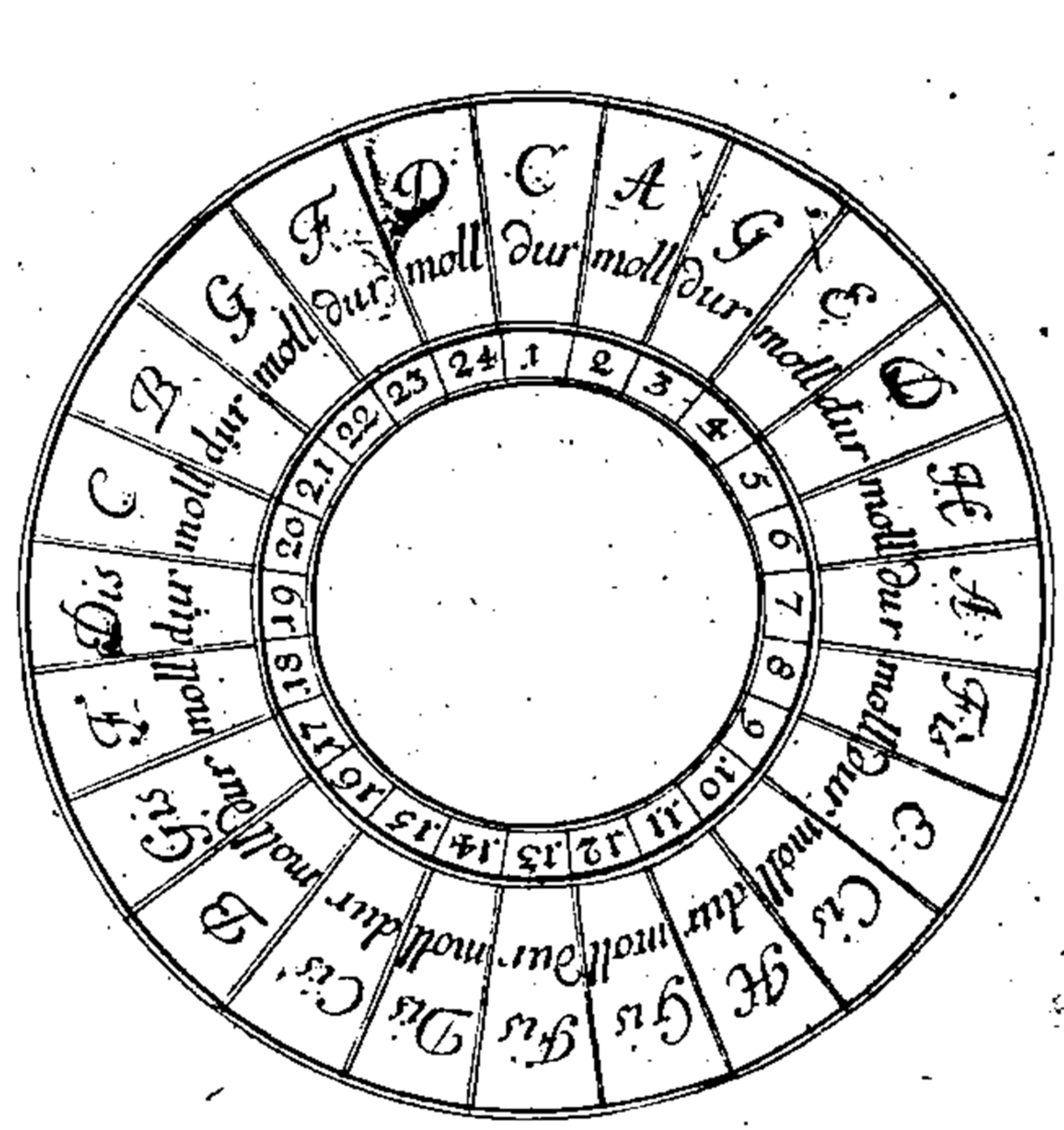}\label{Fig:Heinichen}}\hfill
		\subfigure[][Separation of major and minor chords and their corresponding accidentals]{\includegraphics[width = 0.49\linewidth]{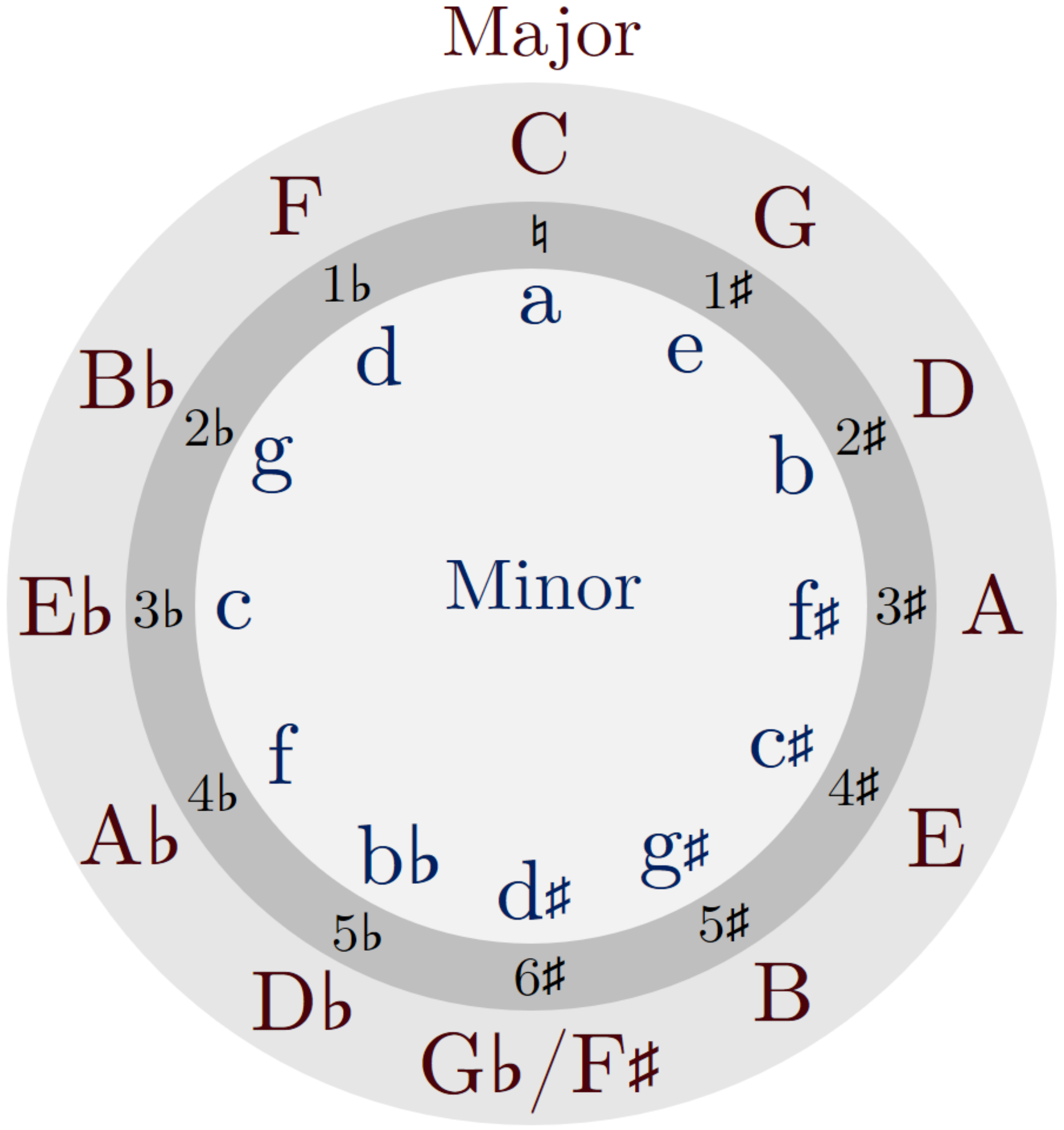}\label{Fig:COFsimple}}
		\caption{The circle of fifths shows the relationships between the different 24 available keys in Western music. \label{cof}}
		\Description{The circle of fifths. Originally published by Heinichen in 1728.}
	\end{figure}

	\mm{
		The \emph{circle of fifths} is a practical concept to explain the geometric structure of the chord relationships between all twelve chromatic pitch classes and is often used by musicians and in music pedagogy~\cite{taylor1989ab} to visualize the relationships between pitch classes. In 1728, Heinichen augmented the circle of fifths to introduce the relationship between major and minor keys, representing the circle of fifths with 24 segments as shown in Figure~\ref{Fig:Heinichen}. Figure \ref{Fig:COFsimple} facilitates the understanding of the parallel minor and major chords while additionally providing the number of accidentals for each key. The distance of the pitch classes in the circle is a measure for the tonal similarity of the keys. For example, if \textit{C~Major} is the tonic, then \textit{F~Major} or \textit{G~Major} are the most similar chords in terms of auditorial perception. 
	}
	
	\subsection{Design Rationale}
	In the glyph's design, we take into consideration multiple musical characteristics which determine the annotated visualization and the used annotation type:
	
	\subsubsection{The Region of Interest}\ \\
	In the design of the harmonic fingerprint, we capture the harmonic variations at a detailed level, while retaining a general overview of the piece. For this reason, we choose single bars as the region of interest for which we automatically create the glyph. Since the tonal center remains the same within a bar, particularly for music with dance-like structure~\cite{taylor1989ab}. Representing harmonic content per bar provides a reasonable compromise between detail and a general overview of the score. 
	
	\subsubsection{Representation of Harmonic Content}\ \\
	To represent the harmonic fingerprint of the bar, we use the root note of the chord within the bar, if it is identifiable. The root note is a common way of describing a chord in music theory, and thus, the glyph uses a description which is familiar to musicians. Other alternatives from music theory would include describing the chord using Roman numerals or \emph{figured bass}~\cite{taylor1989ab2}, but we think that this would be less meaningful to beginners or non-musical experts. The fingerprint allows for distant reading of harmonic relationships in single bars summarizing all notes into a radial chromatic histogram.
	
	\subsubsection{Representation of Melodic Content}\ \\ 
	The glyph captures the note pitches which also form the melodic content of the bar. Since the role of the glyph is to provide a general overview of the harmonic material, we represent this through a radial normalized histogram of the pitch class of notes within the bar. Notes with the same pitch name but in different octaves are grouped in the same histogram bin allowing to identify harmonic relationships across a piece. In this manner, bars which have similar melodic patterns but written in different octaves, or for which there is rhythmic variance, will obtain the same histogram. At an overview level, this is desirable as it captures the broad similarities between bars while keeping close-reading possible through keeping the CMN unchanged. 
	
	\subsubsection{Capturing Harmonic Relationships between Notes}\ \\
	The glyph visualization provide an overview of the harmonic relationship between notes in the bar. This overview should allow musicians to understand at a glance, the presence of consonances or dissonances between notes in the respective bar. Neighboring segments in the fingerprint visualization are the most similar notes in music harmony theory, whereas segments that are placed on the other side of the circle are the most dissimilar relationships. In combination, the reader can efficiently identify the dominant notes for each bar. The circle of fifths metaphor enables us to illustrate the harmonic relationship using a musical concept that is familiar to the musician. 
	
	\subsection{Creating the Glyph}
\begin{figure*}[t]
		\begin{minipage}{\textwidth}
			\subfigure[][A radial representation of how the colors are mapped to the pitch classes according to the circle of fifths.]{\includegraphics[align=c,height=1.55in]{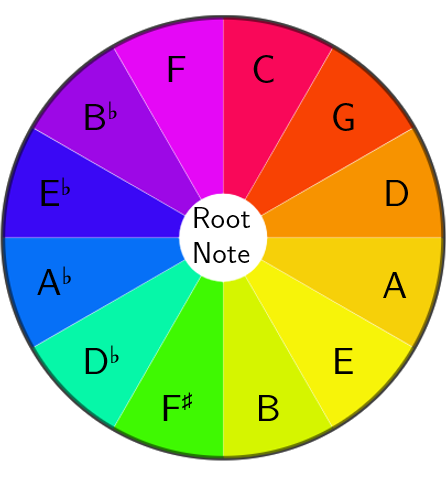}\label{Fig:Colours}}
			\hspace*{-.02in}  
			\includegraphics[align=c,height=0.15in]{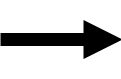}
			\hspace*{-.02in}  
			\subfigure[][Bar 6 bar taken from Chopin's ``Grande Valse Brillante'' using the color encoding from (a) on the  notes.]{\includegraphics[align=c,height=1.55in]{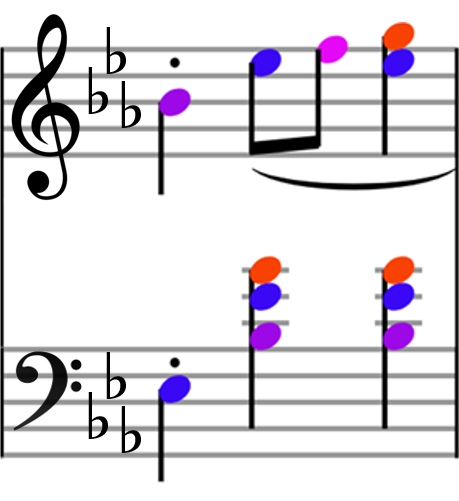}\label{Fig:ExampleScore}}
			\hspace*{-.02in}  
			\includegraphics[align=c,height=0.15in]{img/diag/arrow.png}
			\hspace*{-.02in}  
			\subfigure[][We calculate a colored histogram based on the chromatic pitch classes of all the notes that are present within a bar.]{\includegraphics[align=c,height=1.55in]{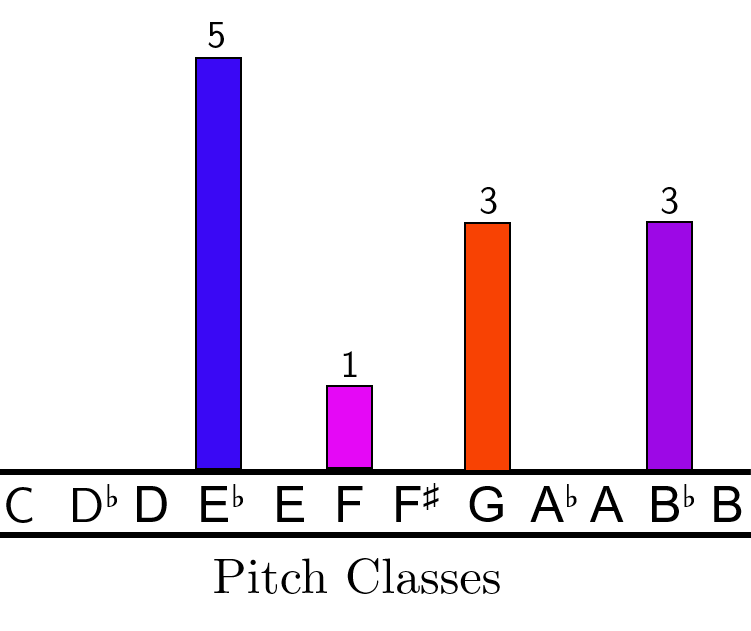}\label{Fig:HistCOF}}
			\hspace*{-.02in}  
			\includegraphics[align=c,height=0.15in]{img/diag/arrow.png}
			\hspace*{-.02in}  
			\subfigure[][The final fingerprint encodes the notes distribution of a bar by segment area size and color.]{\includegraphics[align=c,height=1.55in]{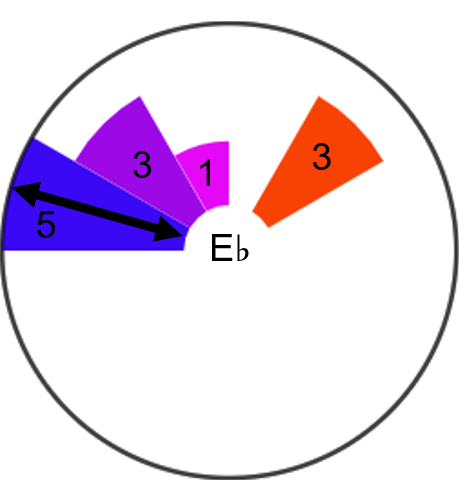}\label{Fig:Glyph}}
		\end{minipage}
		\caption{We use the circle of fifths as a visual metaphor to encode the distribution of notes in a bar (b) by mapping all notes to their corresponding pitch class (c). The amount of notes of each class is then encoded by the size of the area segments as shown in (d). The pitch classes are encoded both using color and the segments position inside the fingerprint (a).}
		\Description{An explanation of the glyph structure}
	\end{figure*}

	We illustrate in Figure~\ref{Fig:Glyph} the specific glyph shape using the 6th bar from Chopin's waltz ``Grande Valse Brillante'' as an example. A histogram of the pitch classes, displayed in Figure~\ref{Fig:HistCOF}, captures the information about the notes forming the melodic and harmonic content. This representation is quite sparse and as a result, would clutter the annotation. We alter the histogram by rearranging the pitch classes radially such that each of the twelve pitch classes from the chromatic scale is displayed by subtending angles of 30\textdegree~(360\textdegree $/$ 12 = 30\textdegree) as shown in Figure~\ref{Fig:Colours}. This allows us to represent the histogram's content more compactly. The number of note occurrences within the bar is encoded by the radius of the segment, resulting in a visualization that is similar to a non-stacked Nightingale Rose Chart~\cite{brasseur_florence_2005}.
	Moreover, rearranging the pitch classes into the circle of fifths format highlights the harmonic relations within the predefined music sheet window. Hence, the harmonic fingerprint can be considered as a statistical overview of the number of notes from a single bar. The reading complexity of the glyph remains unchanged due to its independence of original music document complexity which enables high scalability. 
	
	To further improve the readability of the visual fingerprint, we use the color scale from \citeauthor{ciuha_visualization_2010}, which we show in Figure~\ref{Fig:Colours}.  This color scheme reflects the distance of pitch classes in the color space~\cite{ciuha_visualization_2010}. The visual double encoding of the pitch classes in color and position simplifies the comparison of different fingerprints. 
	
	Lastly, to further emphasize the harmonic content of the bar, we display the root note of the chord formed by the notes in the bar at the center of the glyph as illustrated in Figure~\ref{Fig:Colours}. The resulting shape structure of the glyph is similar to that shown in Figure~\ref{Fig:Glyph}.
	
	\subsection{Adding Harmonic Fingerprints to the Score}
	
	For the scope of this work, we assume that sheet music is readily available in MusicXML file format~\cite{good_musicxml_2001}. MusicXML is an XML-based digital sheet music interchange format designed to provide a universal format for CMN. The use of MusicXML is becoming more widespread and music writing software such as \emph{MuseScore}\footnote{https://musescore.org/en}, \emph{Sibelius}\footnote{https://www.avid.com/sibelius-ultimate} among many others, support MusicXML representations of new works. Likewise, digital libraries such as the Mutopia Project\footnote{http://www.mutopiaproject.org/} provide MusicXML sources for classical works.  
	
	To obtain the statistical data required for the glyph the MusicXML file is pre-processed using the \texttt{\emph{music21}} Python libary~\cite{cuthbert_music21_2010}. Through this library, we compute the number of occurrences of each pitch class, identify the chord and the chord's root note for each bar in the score. 
	
	To display the annotated sheet music with the harmonic fingerprint, we use \emph{OpenSheetMusicDisplay}\footnote{https://github.com/opensheetmusicdisplay/opensheetmusicdisplay}, an open-source application to display MusicXML files in a browser.
	
	Similar to the observations made by \citeauthor{de_prisco_evaluation_2018}, we place the glyphs directly above the bars such that the viewer can intuitively detect the corresponding fingerprint~\cite{de_prisco_evaluation_2018}.  To ensure that the annotations do not overlap with the CMN symbols, we increase the space between each system to provide enough space for the fingerprint annotations. We introduce this additional space through increasing the standard distance between systems in the OpenSheetMusicDisplay application by an amount proportional to the diameter of the glyphs. 
	The allocation of this extra space ensures that both the harmonic fingerprint and the sheet music are legible. 
	
	\section{Evaluation}
	\label{evaluation}
	To evaluate the performance of the fingerprint annotations, we conducted a user study to analyze music theme pattern identification tasks on different sheet music to investigate the users' performance with and without the fingerprint  annotations. We specifically focused in revealing differences between domain experts and non-experts to assess the intuitiveness of our approach as well as identifying tasks that can be supported by the fingerprint visualization introduced in the previous section.

	\subsection{User Study Methodology}
	The assessment of the user study is based on the overall satisfaction reported by the users and their performance in finding the patterns of the provided music sheet sections. 
	
	\begin{figure*}[t]
		\begin{center}
			\includegraphics[width=\textwidth]{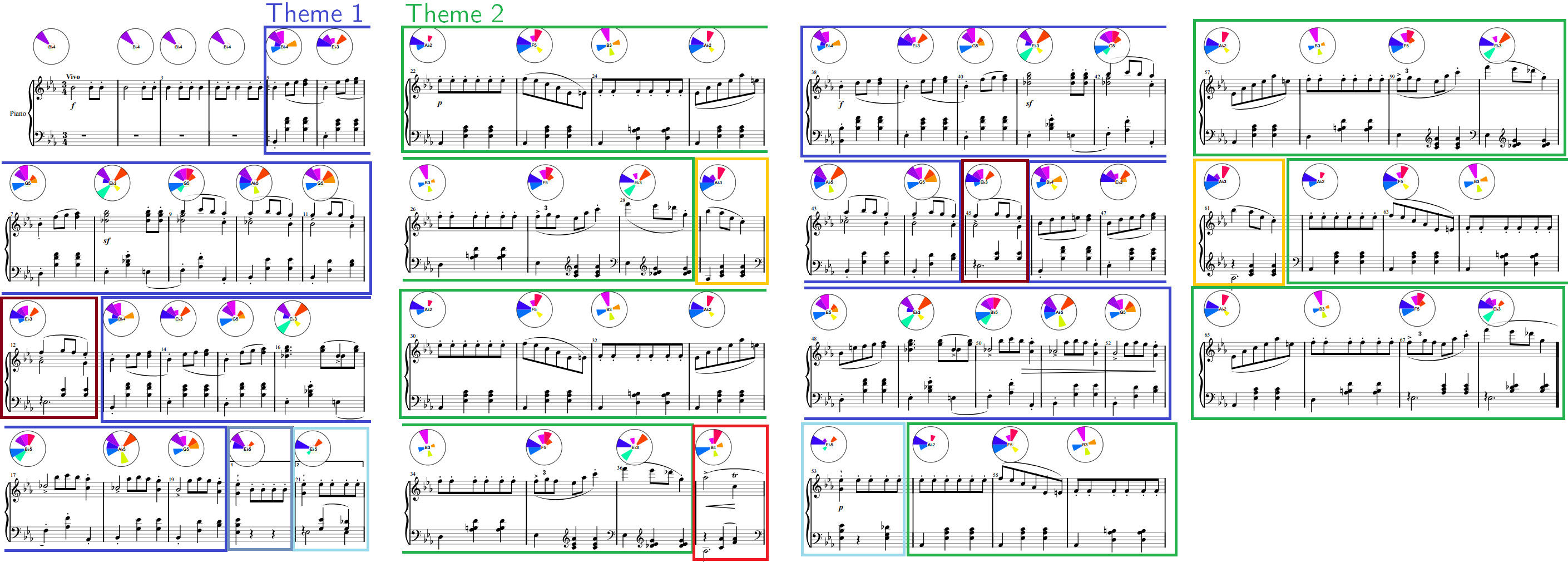}
		\end{center}
		\caption{The first 68 bars of the `Grande Valse Brillante' (forming MS1 of user evaluation). This consists of two major themes which are enclosed here by the blue and green rectangles. Bars 12 and 45, enclosed in yellow and red,  divide the two themes into two parts, with the second part of each theme consisting of some rhythmic variations to the first occurrence of the theme.} 
		\Description{.}
		\label{SM1_a} 
		\vspace{-5pt}
	\end{figure*}

	To perform the evaluation, we conduct a within-subject evaluation, giving each participant two scores, one with and the other without annotations. We instructed the participants to mark any patterns that they can identify on the printed music score, without explicitly telling them to perform harmony analysis. To reduce the potential for bias in the evaluation, we selected the scores such that they are of equal length and complexity. Moreover, we prepared an annotated and a non-annotated version of both scores randomizing the order with which we presented the scores. 
	
	\subsubsection{Dataset and Controls} \ \\
	We applied the fingerprint annotations algorithm to the Fr\'{e}d\'eric Fran\c{c}ois Chopin's ``\textit{Grande Valse Brillante}''~\cite{chopin_imslp_2019_grande_valse_brillante}, for which we processed a ground truth musical harmony analysis beforehand. In total, this waltz has 311 bars of music arranged into seven distinct themes which can be distinguished from the melodic and harmonic variations between the themes. Figure~\ref{SM1_a} shows the first 68 bars of this waltz highlighting the themes. By comparing the annotations obtained from the proposed fingerprint annotation system, it is possible to determine whether the fingerprint annotation system is indeed representing these thematic divisions. 
	
	Since the music score is sufficiently long, we use selected parts in the evaluation as discussed hereunder. We extracted two suitable music-sheet sections which we denote as MS1 and MS2 respectively, with MS1 consisting of the first 68 bars of the waltz while MS2 consists of 64 bars, from bar 69 up to bar 132. Bar 68 has a natural break in the music dividing MS1 and MS2 into two different musical parts. Both MS1 and MS2 consist of recurring patterns: MS1 consists of two major themes as shown in Figure~\ref{SM1_a}, while MS2 consists of a very similar structure but has three different themes instead of two. The full length of the themes varies between 7 or 15 consecutive bars, with the latter containing four-bar sub-patterns. These themes were used as ground truth to estimate the performance of the participants.
	
	\subsubsection{Participants} \ \\
	\mm{
		In total, we evaluated the fingerprint annotation approach with {eight} participants from diverse backgrounds who completed all phases of the user study. We selected the participants by dividing them into two equal-sized groups. 
	}
	
	\mm{
		The first group consists of four domain non-experts (N1--N4) who have little to no music background (mean score of 1.5) and have never performed a harmony analysis (mean score of 1.0). However, these non-experts are experts with a post-graduate degree in the field of data analysis and visualization with a mean age of  {34 $\pm$ 9}.
	}
	
	\mm{
		The second group is represented by people who are either playing instruments daily or have fundamental to solid knowledge about the harmonic relationships in sheet music (mean score of 3.75). Within the scope of this paper, we refer to the second group as our domain experts (E1--E4). These four participants have an intermediate musical knowledge {(mean score of 3.25)}. Three expert participants practice music as their hobby on a daily bases, while the last is singing in a choir but is aware of the harmonic relationship of multiple voices. 
	}
	
	\subsubsection{Study Design} \ \\
	We conducted the study in three phases. In the preliminary phase, we collected general demographic information from the participants, namely their age, gender, level of musical background and familiarity with harmony analysis. To collect the information related to musical background, we asked participants to rate their knowledge on a five-point Likert scale, from \emph{novice} (1) to \emph{expert} (5). Likewise, we asked the participants to rate their familiarity with harmony analysis from \emph{no knowledge} (1) to \emph {very familiar} (5).
	
	In the second phase of the study, we asked the participants to perform the pattern identification task by adding handwritten annotations to highlight patterns on the musical score. The participants carried out the task twice, once with a score containing fingerprints and once with the CMN only. We recorded the amount of time taken by the participants in performing the tasks, advising them that the analysis should not exceed 30 minutes. We deliberately decided to exclude a specific training task to investigate the accessibility to the music document on different knowledge levels without further explanation of the musical scores or the harmonic fingerprint.
	
	The final phase of the study consisted of a short interview to elicit user feedback helping us to answer the research questions stated at the end of the introduction section. We asked participants to receive valuable feedback to gauge their opinion on the usefulness of the harmonic fingerprint and their strategy to fulfill the given pattern identification task.

	\subsection{Use Cases}
	Before presenting the qualitative feedback from the user study interviews, we describe two use cases that are supported by our harmonic fingerprint annotations.
	
	\subsubsection{Identify Octave-Invariant Harmony Relationships} \ \\
	\mm{
		Identifying harmony requires to understand how multiple notes represent a tonal center. This chord depends only on the notes' pitch classes of the chromatic scale and is independent of the octavation. Consequently, the absolute tone pitch does not change the harmony if the pitch class remains unchanged. To identify such similarities, a reader has to identify the line for each note and apply the accidentals of the current key to extract the underlying chord. This is a tedious task if the number of notes increases. 
	}
	\mm{
		For example, if the note C is available three times simultaneously (e.g., C2, C4, C5) in addition to other notes, then the chord identification task becomes more complex. The fingerprint annotation simply merges all pitch class occurrences making it easier for analysts to focus on the composition of pitch classes to decide which harmonic aspects are most dominant in a bar and define the chord.
	}
	
	\subsubsection{Musical Theme Extraction} \ \\
	\mm{
		Depending on the composition, the complexity of the CMN makes it difficult to find recurring musical themes in a score. By using a unique color scale to represent each pitch class, the annotated fingerprint enables readers to efficiently skim a document for similarity. Since color is easier to differentiate then black and white, readers can visually filter for similar harmony patterns. Moreover, the layout of the CMN symbols, such as stem direction, note duration, or note divisions have no visible influence on the fingerprint and provide a trustworthy foundation to identify harmonic differences and commonalities. Figure~\ref{SM1_a} displays how music themes can be identified by comparing the fingerprint visualizations across a score. Due to distant reading, even small differences of varying pitch classes can be detected. By keeping the CMN, close-reading is still possible, if required to compare rhythmic details within the bars. In addition, the fingerprint enables to identify the most frequent note in each bar at a glance.
	}

	\begin{figure*}[t]
		\begin{center}
			\includegraphics[width=\textwidth]{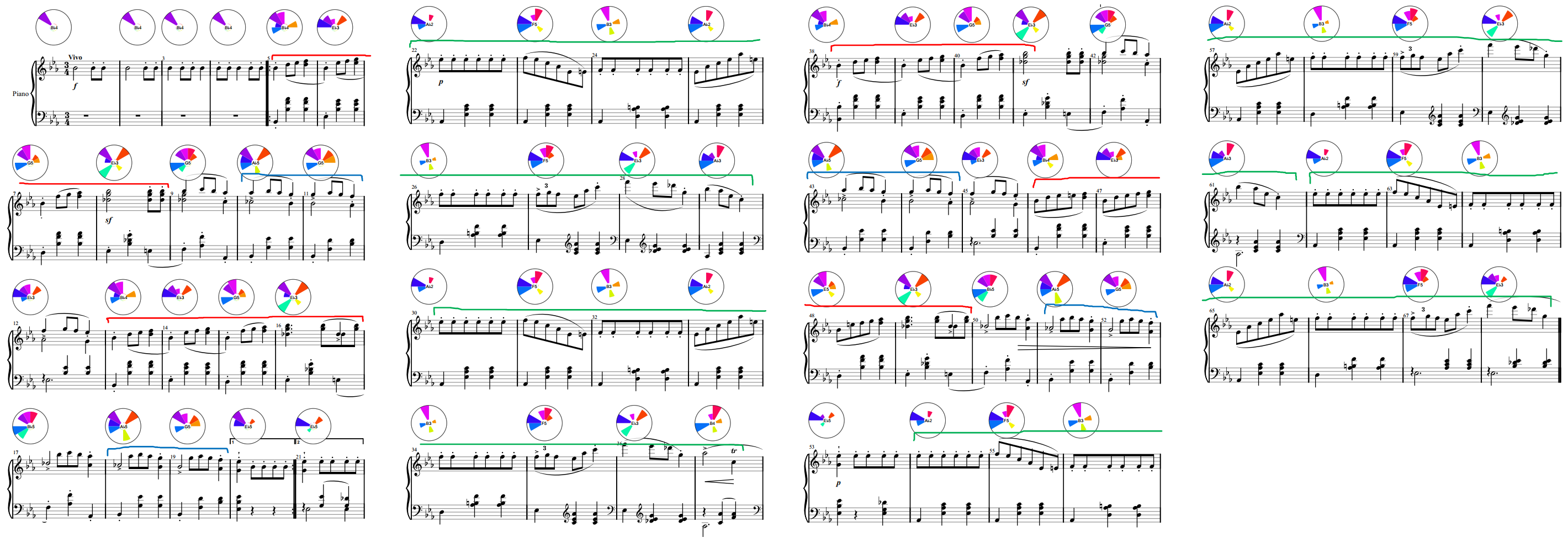}
		\end{center}
		\caption{We told the participants to add handwritten annotations to the printed sheet music. The figure shows how E3 highlighted different patterns. E3 was provided with MS1 including the fingerprint visualization. She highlighted the first half of \textit{Theme 1} in red and its last two bars using blue color. She identified \textit{Theme 2} completely by marking it in green.} 
		\Description{Participants added handwritten annotations to the sheet music to mark the found themes or patterns. The performance of the participants could be compared by counting the patterns they found in the printed music document.}
		\label{SM1_E3} 
	\end{figure*}

	\subsection{User Feedback}
	\label{results}
	Figure \ref{SM1_a} shows the annotation results for the bars 5--68 of Chopin's ``Grande Valse Brillante'' are shown superimposed on the ground truth thematic analysis of this waltz. The second occurrence of theme 1 has similar fingerprints as the first occurrence. There are some rhythmic differences in the melody line of the second occurrence which are not reflected in the fingerprint visualization. 
	The fingerprints for the second theme are considerably different from that of the first theme, highlighting the shift in the key.
	Recurring themes have very similar fingerprints, as expected.

	To understand the effect of the harmonic fingerprint glyph for music readers that are on different music expertise levels, we report the qualitative results and elicited feedback on the usefulness of our visualization from the participants. Based on this feedback and the users' performance in finding the themes we will address the research questions Q1, Q2, and Q3 from the introduction. We compared the differences between the participants' annotations and the ground truth of the harmonic patterns that are included in the datasets. We highlight the participants' statements by using italic and double quotation marks in the following paragraphs.

	\subsubsection{Usefulness of the Fingerprint} \ \\  
	Except for N3, all participants stated that they would like to use the fingerprint annotations in the future in pattern identification tasks having different reasons. For example, N1 ``\textit{used the fingerprint as an anchor and to keep the overview, [but] got lost in the CMN because of the notes' optical similarity.}'' N2 explained that by using the glyph, she could ``\textit{compare images}'' to identify the pattern because it is easier since the complexity of the CMN makes it hard to identify differences. First, N3 performed the pattern identification task without the fingerprint and was confused about the fingerprint because the ``\textit{notes where easier to compare}'' and the fingerprint seemed to not match with the CMN. For N4, it was ``\textit{much easier with the fingerprint [...] to see patterns based on the color and location [...] than trying to scan and compare the notes.}''
	
	E1 argued that ``\textit{to practice a piano piece, it might be helpful to spot parts which are equal or have small differences.}''
	E3 found it ``\textit{an interesting way of putting a visual aspect to the combination of notes that are indicated within the music score.}''
	E4 stated that it is  ``\textit{easier to find the patterns with the visualization,}'' but was unsure whether she might have missed some important information because she did not look further into the CMN. E4 could ``\textit{recognize how the piece develops}'' and ``\textit{obtain an overview over the piece easier.}''
	
	\subsubsection{Pattern Identification Strategy} \ \\ %
	Depending on whether the fingerprint was annotated to the sheet, the different user groups applied different strategies to identify repeating patterns. All non-experts, except N3, preferred the annotated condition since they found it easier to find matchings. N1 ``identified differences and similarities by comparing first the presence of the glyph's segments and afterward its size without looking at the CMN anymore if the glyphs matched''. N4 stated: ``\textit{I started with the first icon [...] to find another that matched until the next.}'' N2 and N4 only highlighted the patterns by adding handwritten annotations by grouping the fingerprints instead of using the CMN. 
	N1 ``\textit{identified differences and similarities by comparing first the presence of the glyph's segments and afterward its size without looking at the CMN anymore if the glyphs matched.}'' N1 and N3 incrementally summarized subpatterns without identifying a theme in full. 
	
	In the first round, E1 was provided with the non-annotated document where he textually added the chords symbol to the first 12 bars. E1 continued by highlighting the melodies throughout the document without looking at the harmony any further. E1 found the fingerprint helpful ``\textit{to spot bars which are the same}'' and was ``\textit{gathering information by switching focus between glyphs and [the] music sheet.}'' 
	At first, E2 ``\textit{skimmed through the whole piece to gain an overall picture of the piece [and then] moved on to analyzing bar by bar [to] compare phrases to each other to find where repetitions occurred.}'' Figure~\ref{SM1_E3} displays E3's handwritten annotations: E3 divided \textit{Theme 1} in MS1 as two subpatterns, excluding bars 9, 12, and 17, and completely identified \textit{Theme 2}. 
	
	To find similarities, E3 ``\textit{tried to merge the harmonic fingerprint glyph with the melodic patterns to identify the sequences present}'' and ``\textit{followed the labeling of the chords as well as the note patterns [...] at a different pitch level.}'' To identify recurring ``\textit{melodic patterns}'' in MS2, E3 applied the same strategy as for MS1 with the glyphs by ``\textit{marking the tonal and real sequences [...] through the identification of the patterns where the note intervals are different [...], but the pitch is moving in the same direction.}''
	
	\subsubsection{Challenges} \ \\
	The participants faced various challenges in the execution of the assigned task. For example, N1 stated: ``\textit{I had to go back to the bars that I wanted to compare more often because I couldn't remember all the details as good as with the fingerprint.}'' 
	N1 argued that he ``\textit{found differences in the lower system, while the upper system did not change within a bar sequence.}'' 
	N1 mainly annotated shorter patterns and could not identify a complete theme. Similarly, N3 did not find complete harmonic theme patterns without the fingerprint but often highlighted subpatterns of two subsequent bars. N1 also gave the feedback that ``\textit{the fingerprint is easier to remember how they look compared to the original music sheet, especially if patterns are farther away.}''
	
	N2 found it difficult to ``\textit{identify patterns that have optics jumps [because of] new rows continuing on the next page.}'' N3 was the only participant, who considered the fingerprint to be confusing since it indicated similarity whereas the music sheet did not reflect this. Hence, N3 found the glyph visualization misleading in finding exact patterns. N3, who was provided with MS2 without the fingerprint, indicated that it is ``\textit{difficult to see exactly on which line the notes are placed and if they are similar to the others.}'' 
	In the second round, N4 had to find the patterns without the glyphs. Due to their absence, N4 perceived it to be ``\textit{more tedious to look more closely at the single notes,}'' especially if the patterns broke across multiple systems.
	
	For E1, the most challenging aspect of the analysis task was in ``\textit{reading the sheet and to decide which overall chord it is [because] it is time-consuming.}'' One of the most challenging aspects for E2 and E4 was ``\textit{to analyze the piece and find similarities [...] without having heard the [...] music being played, as this would make such repetitions extremely evident.}'' E3 ``\textit{was not quite sure what the colors within the harmonic fingerprint glyph represented.}'' Due to the bar-based distribution of the fingerprints, E4 declared to be ``\textit{too focused on single bars at the beginning}'' requiring more time to find larger patterns in the sheet.

	\section{Discussion and Lessons Learned}
	\mm{
		The qualitative user study that we conducted to analyze the performance of our fingerprint annotation revealed both benefits and drawbacks compared to the CMN only. Except for N3, all non-experts could identify many parts or full patterns of the themes without the need for looking at the CMN details. Thus, the fingerprints can facilitate the access to sheet music for people that have little to no knowledge about music theory or harmony rules because it encodes salient harmonical characteristics (Q1). In this way, novices can become more motivated to look at sheet music to understand the relationship between the notes in a bar and the annotated visualization. N2 and N4 found the visualization to be visually compelling which increased their interest to look through the music sheet (Q2). Novices and music learners often become overwhelmed by the complexity of the CMN and give up early due to the steep learning curve in understanding the CMN. 
	}
	
	\subsection{Educational Aspects}
	\mm{
		Since the circle of fifths is the foundation of harmonic relationships in Western music, the fingerprint is a suitable method to teach learners about the harmonic structure in sheet music. It provides a quick overview of bars and facilitates the differentiation of bars that appear in different sections of a music sheet. Mainly, the fingerprint allows one to efficiently identify those bars which have the same visual melodic pattern, but which are written at a different pitch. Consequently, these bars can be distinguished from bars that are only transposed by octaves without the need for close reading of the single notes in CMN.  
		To check rhythmic differences, users are still able to look at the original notation, since we did not change the CMN in the processing step of the fingerprint annotation algorithm.
		Participants that performed the pattern search task without the fingerprint primarily focused not only on the pitch class but on the rhythmic characteristics. Since the fingerprint only encodes the harmonic characteristics, it serves as a natural filter of the rhythmic aspects which are, besides the note pitches, the most dominant optical features in sheet music.
	}
	
	\subsection{Staff Separation}
	\mm {
		An interesting finding was that non-expert users who first worked through the task having the fingerprints attached to the sheet, tended to combine the upper and lower staves in the second condition, even when the fingerprint was not given any more (Q2). Due to typical structures in the different staves, Two participants of each group identified the patterns without combining the staves. Therefore they did not differentiate between these parts, even when combined the resulting harmony is different. 
		They analyzed the upper and the lower stave separately to find harmonic similarities when the fingerprint was not present by starting with patterns with a length of single bars and extended them until they identified differences. The remaining study participants recognized the fingerprint as a summary of the bars and seldom looked further into the details of the CMN (Q2). We call these contrasting strategies top-down (with fingerprints), and bottom-up (no fingerprints) analysis approaches. The top-down approach was mainly applied by N1--N4 by including the fingerprint when adding handwritten annotations to the printed music sheet.
	}
	
	\subsection{Harmonic Invariance}
	\mm {
		In the third theme of MS2, all notes from bars 119--124 are transposed by an octave higher in bars 127--132. To identify this similarity, the reader has to identify the single notes to understand that harmony is similar. Both N2 and N3 were provided with MS2 without the fingerprint and could not identify this harmonic similarity. In comparison, N1 and N4 used the harmonic fingerprint to see the similarities without knowing the underlying relationships. With having the domain knowledge, E2 and E3 were able to detect this harmonic similarity even without being provided with the visual annotation. E4 marked this harmonic similarity by highlighting the fingerprints whereas E1 highlighted the pattern in the CMN. Hence, we assume that when readers want to find harmonic patterns of all simultaneous notes, the fingerprint is supportive of extracting such similarities more efficiently. 
	}
	
	\mm{
		Another drawback that we extracted from the feedback of E1--E3 is that they tend to ignore the fingerprint in favor of the CMN because of its familiarity (Q3). For this user group, the fingerprint representations are less intuitive than the CMN. Since we did not explain the design decisions of the visualization to the participants, we elicited that users who do not try to understand the detailed harmonic relationship trust the fingerprint without questioning. One music student replied that she was not sure about the color coding in the glyph. Nonetheless, she still found it a promising way to put a visual aspect of the combination of notes that are indicated within the music score.
	}

	\subsection{Limitations and Future Work} \ \\
	The fingerprint visualization that we designed to reveal harmonic relationships in sheet music has proved to be helpful to readers in some scenarios including harmonic pattern identification tasks. Nevertheless, we identified some issues in the current design. 
	
	First, only one color map can be selected to encode the pitch class that is difficult or impossible to read for people who are visually impaired. Secondly, we currently combine all notes in a bar to extract the root of the dominant chord. Therefore, the fingerprints certainly show the exact notes histogram for each bar, which is equally considered in calculating the tonal center for the bars. Consequently, the root note that is displayed at the center of the fingerprint is sometimes not correct. For example, due to the tonal relationship of keys, the root note of the other chord mode of the relative key, or even a single note of a melody sequence that is not even part of the actual chord is selected as the root by \texttt{music21}. In the future, we plan to improve the accuracy by weighting the notes by their duration and excluding underrepresented notes from the calculation of the root note. Similarly, we want to additionally reflect the dominance of notes by their duration in the fingerprint. For example, if a note is played only once but has a long duration, then it is currently underrepresented by the harmonic fingerprint and should be weighted according to its tone length.
	
	Both musicians and music analysts are not only interested in the distribution of notes in a bar but want to know if the tonal center is a minor or major chord. We will include this information in the fingerprint to save readers the time in extracting this information from the glyph visualization which is currently required. We think that it can be an interesting experiment to see how differences in music styles are reflected in the harmonic fingerprint that can visualize the salient characteristics of music. 
	
	Based on the qualitative user feedback, we elicited valuable feedback regarding the information that is encoded in the fingerprint visualization. Music readers who have no understanding about the underlying music theory require an introduction to the fingerprint because one may assume that it encodes all musical information. Afterward, they can better estimate whether the visualization is suitable for a given task. This has led to confusion for some of our non-expert paricipants who did not trust the visualization after identifying presumed differences due to the harmonic invariance between the fingerprint and the CMN. 
	When conducting a quantitative evaluation we recommend to explain the visualization to the participants in advance to enhance the performance in harmony analysis tasks by improving its understanding. It will be useful to investigate the suitability for different tasks to elicit in which situation the fingerprint supports the reader and when it is better to only show the CMN. 
	
	Eventually, contrary to our expectations, understanding the meaning of the fingerprint visualization is not readily intuitive even for domain experts. This can be a result of the experts' familiarity with the CMN, but also on their definition of a pattern, which was not specified in the preliminary phase of the user study. We aim to conduct another survey which will include a detailed explanation of the fingerprint to the users, to see whether the performance increases in different music analysis tasks, if the reader is aware of the encoded information.

	\section{Conclusion}
	We introduced a visualization method to encode salient harmonic characteristics as a fingerprint. For the design of the \textit{harmonic fingerprint glyphs},~\footnote{Our system is available under \url{https://musicvis.dbvis.de/app/fingerprint}} we exploit the circle of fifths, which is the foundation for harmony relationships in Western music, as a visual metaphor to augment digital music sheet with additional harmonic information. To analyze the usefulness of our approach, we conducted a qualitative user study with four domain experts and four participants without a musical background. The evaluation revealed a potential for identification of recurring harmony progressions and the understanding of the underlying harmonic structures in sheet music, even if the optical appearance suggests dissimilarity. Moreover, our method is suitable to support distant and close reading in sheet music exposing harmonic relationships on a rather abstract level, while keeping the original music notation unchanged.
	As a consequence, readers can view sheet music on different levels of detail. In the future, we aim to enhance our approach by integrating user feedback and by setting up a visual analysis system for music scores. Thus, we further aim at combining close and distant reading into an interactive visual analysis system. Consequently, this web application will enable fast access to suitable music analysis tools for any sheet music available in MusicXML.

	\bibliographystyle{ACM-Reference-Format}
	\bibliography{references}
	
\end{document}